\documentclass[11pt, prd,nofootinbib,preprint,superscriptaddress]{revtex4}
\pdfoutput=1
\usepackage[T1]{fontenc}
\usepackage{amsmath,amssymb}
\usepackage{epsfig}
\usepackage{subfigure}
\usepackage{float}
\usepackage{graphicx}
\usepackage{slashed}
\usepackage{multirow}
\usepackage[colorlinks,citecolor=blue]{hyperref}
\usepackage{pdfpages}
\usepackage{color}
\usepackage{booktabs}
\usepackage{comment}
\usepackage{ulem}
\usepackage{soul}
\usepackage[fleqn]{mathtools}

\definecolor{orange}{rgb}{1,0.5,0}

\DeclarePairedDelimiterX\braket[2]{\langle}{\rangle}{#1\,\delimsize\vert\,\mathopen{}#2}

\def\mET{\slashed{E}_T}

\begin{document}

\title{{Unveiling the CP-odd Higgs in a Generalized 2HDM at a Muon Collider}}

        \author{Nandini Das}
	\email{nandinidas.rs@gmail.com}
	\affiliation{School of Physical Sciences, Indian Association for the Cultivation of Science, 2A $\&$ 2B, Raja S.C. Mullick Road, Kolkata 700032, India}
 
	 \author{Nivedita Ghosh}
  \email{niveditag@iisc.ac.in}
\affiliation{Centre for High Energy Physics, Indian Institute of Science, 2A $\&$ 2B, Raja S.C. Mullick Road, Bengaluru, 560012, Karnataka, India}

\begin{abstract}
We revisit the generalized two-Higgs-doublet model (2HDM) with a \textit{minimally perturbed} lepton specific (Type-X) Yukawa sector in context of the muon collider. The model offers a large region of parameter space where it can provide a solution to the observed muon ($g-2$) excess. Specifically the low mass CP odd scalar of this model plays an important role to fit this anomaly. Interestingly, the coupling of the pseudo scalar with the muons is proportional to $\tan \beta$ (the ratio of VEVs of the Higgs doublets) which dominantly controls the one loop and two loop contributions of the muon magnetic moment, therefore it can be probed directly at a muon collider with greater sensitivity in comparison to Large Hadron Collider (LHC). Primarily focusing on a parameter space where muon $g-2$ anomaly is satisfied, we explore the available theoretical and experimental constraints arising from vacuum stability, unitarity, lepton falvour violation (LFV), electroweak precision constraint, B-meson decay and collider experiments on the model. In the parameter space where all these constraints are satisfied,  we propose to study  the production of the light pseudo-scalar $A$ in association with a photon at a $3$ TeV
muon collider. Subsequent decay of $A$
into a $\tau$ pair will produce a striking $l^+ l^- \gamma$ plus missing energy signature.

\end{abstract}

\setcounter{footnote}{0}  
\renewcommand{\thefootnote}{\arabic{footnote}}  

\maketitle

\section{Introduction} \label{sec:intro}
Muon ($g-2$) anomaly~\cite{Muong-2:2023cdq} has been one of the long-standing motivations for the particle physicists to look for BSM physics scenarios since last two decades. The latest measurement of anomalous magnetic moment by "MUON G-2" collaboration
at Fermi National Laboratory (FNAL) combined with the E989 experiment at the Brookhaven National Laboratory (BNL) shows a 5.1$\sigma$ deviation from its Standard Model prediction. In order to address this discrepancy, many BSM scenarios~\cite{Chen:2021jok,Iguro:2019sly,Arora:2022hza, PhysRevD.100.015024} have been widely studied in the existing literature. However, the enhanced coupling of muon with non-standard particles (beyond Standard Model particles) which is necessary to explain the muon anomaly in general, has not been extensively probed in colliders yet~\cite{Arakawa:2022mkr,Yin:2020afe}. This coupling can be directly probed in the futuristic muon collider. We here focus to explore such a "muon-philic" scenario.

The proposal of a high-energy muon collider by the International Muon Collider Collaboration (IMCC) is a very  important and interesting development in recent time ~\cite{IMCC,Schulte:2019bdl,Delahaye:2019omf,Schulte:2022brl}. The novelties of a muon collider lie in several facts. Firstly, in the energy frontier, lepton colliders are more efficient in the energy usage of particle collisions compared to the hadron colliders where only a fraction of energy can be used in collision due to the composite nature of protons. Secondly, the hadron
collider is challenged by a noisy environment due to pervasive hadronic activity and
smearing effects from the parton distribution functions (PDFs). All such effects make any precision study very difficult. On the other hand, a muon collider can achieve a very high-energy muon beam due to relatively low synchrotron radiation compared to a $e^+e^-$ collider. The muon, being an elementary particle, can therefore be lucrative in returning high center-of-mass energies in hard collisions along with
a very little energy spread due to the suppressed radiative effects of bremsstrahlung~\cite{Chen:1993dba,Barklow:2023iav}. Therefore, it is safe to say that the muon collider can be a very elegant choice for the coming ages which can provide the advantages of both pp and $e^+e^-$ colliders offering the benefits of high energy and high precision~\cite{Costantini:2020stv,Han:2020uid,Han:2021kes,AlAli:2021let,Accettura:2023ked,MuonCollider:2022xlm}. The energy and luminosity of the upcoming muon collider are not yet finalized. However, there is a proposal to run at 1 $\rm ab^{-1}$ luminosity for a 3 TeV center of mass (c.o.m) energy and 10 $\rm ab^{-1}$ luminosity for a 10 TeV machine~\cite{Costantini:2020stv,Han:2020uid,InternationalMuonCollider:2024dlm}. As the luminosity of the 3 TeV machine is comparable to the 14 TeV HL-LHC luminosity, it is expected that the early stages of the muon collider would take a crucial role in identifying NP signals that LHC might not be able to probe even with its high luminosity option. These aspects make muon collider an impeccable option to search for new physics (NP) scenarios.

In this work, we consider a generalized 2HDM~\cite{Mahmoudi:2009zx,Diaz-Cruz:2010xmg,Bai:2012ex}, with a minimally perturbed Type-X Yuwaka sector~\cite{Crivellin:2015hha}. The non-standard scalars i.e. the charged Higgs and the neutral scalars of this model contribute to the muon magnetic moment via one loop and two loop diagrams. Particularly the pseudoscalar can give dominant contribution in the magnetic moment of muon in its low mass region. The pseudoscalar-muon-muon coupling which dictates the contribution of the diagrams in muon anomaly, can be probed  at muon collider. Though the study of generalized 2HDM in the context of LHC has been studied in great detail~\cite{Atwood:1996vj,Diaz:2000cm,Arhrib:2011wc,Chakrabarty:2017qkh,Iguro:2019sly,Cao:2009as,Chun:2015hsa,Wang:2018hnw,Primulando:2019ydt,Chakrabarty:2021ztf,Dey:2021pyn,Dey:2021alu,Moretti:2022fot,Blanke:2022kpi,Mukhopadhyaya:2023akv,Iguro:2023tbk}, this model with emphasis on this contributory coupling to muon $(g-2)$ anomaly  has not yet been studied at the muon collider ~\cite{Barger:2013ofa,Hashemi:2013sja,Chakrabarty:2014pja,Han:2022edd,Ouazghour:2023plc}. Here we intend to probe a pseudoscalar of $30-50$ GeV mass range in the context of this model at muon collider. After finding a suitable region of parameter space
where along with the muon anomaly and LFV constraints, theoretical constraints
coming from perturbativity, unitarity, vacuum stability,  constraints coming from oblique parameters, B physics and collider experiments are also obeyed, we explore the possibility of probing a pseudoscalar at a 3 TeV muon collider in $\ell^+\ell^{'-}\gamma+\mET$ final state where $\ell,\ell^{'}=e,\mu$. This channel serves as a complementary channel to look for the light pseudoscalar at the LHC. The reason behind this complementarity comes from the Yukawa structure of the model. To satisfy the anomolous magnetic moment of muon, it is observed that moderate to high $\rm\tan\beta$ (ratio of the vacuum expectation values of the two Higgs doublets) is preferred which enhances the lepton Yukawa coupling with the pseudoscalar and simultaneously reduces the same for the quark Yukawa. As a result, we find out that even at HL-LHC, this signal cannot be probed even with high luminosity, whereas at a 3 TeV muon collider, with merely 1 $\rm ab^{-1}$ of luminosity, an ample amount of parameter space of this model can be easily probed with significance $\gtrsim$ 4$\sigma$ making it a more economic choice.

The paper is organized as follows: in section~\ref{sec:model}, we briefly describe the model. In section~\ref{sec:muong-2}, we  discuss the muon anomaly in context of the model and present the allowed parameter space. In section~\ref{sec:cons}, we discuss all possible theoretical and experimental constraints of the model. In section~\ref{sec:coll}, we present a distinct collider signature of the model and provide an explanation for our preference of muon collider over LHC. Finally, we conclude in section~\ref{sec:conc}.

\section{Two Higgs Doublet model} \label{sec:model}
In this section, we briefly discuss the model of our interest~\cite{Primulando:2016eod,Primulando:2019ydt}. For a overview of 2HDM, we refer the readers to Ref.~\cite{Branco:2011iw}. The most general potential containing two $SU(2)_L$ doublet Higgs bosons can be written as 
\begin{eqnarray}
    V(\Phi_1, \Phi_2)&=& m^2_{11} (\Phi_1^\dagger \Phi_1)+m^2_{22} (\Phi_2^\dagger \Phi_2) -[m^2_{12} (\Phi_1^\dagger \Phi_2)+\rm{H.C.}] \nonumber \\
    &+&\frac{1}{2} \lambda_1 (\Phi_1^\dagger \Phi_1)^2 + \frac{1}{2} \lambda_2 (\Phi_2^\dagger \Phi_2)^2
     + \lambda_3 (\Phi_1^\dagger \Phi_1) (\Phi_2^\dagger \Phi_2) \nonumber \\ 
     &+&\lambda_4 (\Phi_1^\dagger \Phi_2) (\Phi_2^\dagger \Phi_1) +\Big\{\frac{1}{2} \lambda_5 (\Phi_1^\dagger \Phi_2)^2 + [\lambda_6 (\Phi_1^\dagger \Phi_1)
     \nonumber \\
     &+& \lambda_7 (\Phi_2^\dagger \Phi_2)]
  (\Phi_1^\dagger \Phi_2) +\rm {H.C.}\Big\}
 \label{eq:potential}
    \end{eqnarray}
    where H.C. stands for the hermitian conjugate of the corresponding term. After electroweak symmetry breaking, the two scalar doublets $\Phi_1$ and $\Phi_2$ can be expanded around the vacuum expectation values(vevs) as 
    \begin{eqnarray}
      \Phi_1 &=& \begin{pmatrix}
      \phi^{+}_1 \\
      \frac{1}{\sqrt{2}}(\rho_1 +v_1 + i \eta_1)
          \end{pmatrix},\\ \nonumber
           \Phi_2 &=& \begin{pmatrix}
      \phi^{+}_2 \\
      \frac{1}{\sqrt{2}}(\rho_2 +v_2 + i \eta_2)
          \end{pmatrix}.\\ \nonumber
    \label{eq:Higgs}      
    \end{eqnarray}
   The ratio of the two vevs is parametrized as $\tan\beta =\frac{v_2}{v_1}$, which plays a key role in the analysis. The singly charged scalars can be written as a linear combination of the following two mass eigenstates, a charged Goldstone boson $G^{\pm}$ and a physical charged Higgs scalar $H^{\pm}$. Similarly, the gauge eigenstates of CP odd neutral scalars can be expressed as a linear combination of  $G_0$, a massless CP odd Goldstone and $A$, a physical massive CP odd scalar.  The gauge eigenstates of charged scalar and
CP odd scalars in terms of mass eigenstates can be written as
   \begin{eqnarray}
       \begin{pmatrix}
           \phi_1^{\pm}\\
           \phi_2^{\pm}
       \end{pmatrix}
       &=& \begin{pmatrix}
           \rm{cos\beta} & \rm{sin\beta} \\
           \rm{sin\beta} & -\rm{cos\beta} \\
           \end{pmatrix}
           \begin{pmatrix}
               G^{\pm} \\
               H^{\pm}
           \end{pmatrix}
    \label{eq:chargedh}       
   \end{eqnarray}
   \begin{eqnarray}
       \begin{pmatrix}
           \eta_1\\
           \eta_2
       \end{pmatrix}
       &=& \begin{pmatrix}
           \rm{cos\beta} & \rm{sin\beta} \\
           \rm{sin\beta} & -\rm{cos\beta} \\
           \end{pmatrix}
           \begin{pmatrix}
               G_0 \\
               A
           \end{pmatrix}
    \label{eq:neutralh}       
   \end{eqnarray}
    
    The CP even gauge eigenstates can be written as 
    \begin{eqnarray}
        \begin{pmatrix}
            \rho_1\\
            \rho_2
        \end{pmatrix}&=& 
        \begin{pmatrix}
                -\rm{sin\alpha} & \rm{cos\alpha}\\
                \rm{cos\alpha}  &  \rm{sin\alpha}
        \end{pmatrix}
        \begin{pmatrix}
            h\\
            H
        \end{pmatrix}
    \label{eq:cpeven}    
    \end{eqnarray}
    \newline
    In the general 2HDM, where no $Z_2$ symmetry is imposed on the Lagrangian, we can write the Yukawa terms of the Lagrangian as
    \begin{eqnarray}
        -\mathcal{L}_{\rm{Yukawa}} &=& \Bar{Q}_L (Y^d_1 \Phi_1+Y^d_2 \Phi_2) d_R + \Bar{Q}_L (Y^u_1 \Tilde{\Phi}_1+Y^u_2 \Tilde{\Phi}_2) u_R \nonumber \\
        &+&
        \Bar{L}_L (Y^l_1 \Phi_1+ Y^l_2 \Phi_2)e_R +H.C. 
        \label{eq:yuk0}
        \end{eqnarray}
        where $Y_{1,2}^{u,d,l}$ are Yukawa matrices and $\Tilde{\Phi}_{i}$ is defined as $ \Tilde{\Phi}_{i}= i\sigma_2 \Phi_i^\star$.

  $Y^i_1$ and $Y^i_2$ in the above equation can not be diagonalized simultaneously without assuming any particular relation. We follow the prescription of~\cite{Crivellin:2015hha} and choose to  
        diagonalize $Y^u_2$,$Y^d_2$ and $Y^l_1$ matrices, while $Y^u_1$,$Y^d_1$ and $Y^l_2$ remains non-diagonal, giving rise to the tree-level Flavor-changing-neutral current (FCNC) in the Yukawa sector. The Yukawa lagrangian for the neutral scalars can be written as
        \begin{eqnarray}
            -\mathcal{L}_{\rm{Yukawa}}&=& \Bar{u}_L \Big[ \Big( \frac{c_\alpha m^u}{v s_{\beta}}-\frac{c_{\beta -\alpha } \Sigma^u}{\sqrt{2} s_\beta}\Big) h+ \Big( \frac{s_\alpha m^u}{v s_{\beta}}+\frac{s_{\beta -\alpha } \Sigma^u}{\sqrt{2} s_\beta}\Big) H\Big] u_R \nonumber \\ 
            &&+
            \Bar{d}_L \Big[ \Big( \frac{c_\alpha m^d}{v s_{\beta}}-\frac{c_{\beta -\alpha } \Sigma^d}{\sqrt{2} s_\beta}\Big) h +\Big( \frac{s_\alpha m^d}{v s_{\beta}}+\frac{s_{\beta -\alpha } \Sigma^d}{\sqrt{2} s_\beta}\Big) H\Big] d_R\nonumber \\
            &&+
\Bar{e}_L \Big[ \Big( -\frac{s_\alpha m^l}{v c_{\beta}}+\frac{c_{\beta -\alpha } \Sigma^l}{\sqrt{2} c_\beta}\Big) h+ \Big( \frac{c_\alpha m^l}{v c_{\beta}}-\frac{s_{\beta -\alpha } \Sigma^l}{\sqrt{2} c_\beta}\Big) H\Big] e_R \nonumber \\
&&+i \Big[ \Bar{u}_L \Big( \frac{m^u}{v t_\beta}-\frac{\Sigma^u}{\sqrt{2} s_\beta} \Big) u_R+\Bar{d}_L \Big( -\frac{m^d}{v t_\beta}+\frac{\Sigma^d}{\sqrt{2} s_\beta} \Big) d_R\nonumber \\
&&+ \Bar{e}_L \Big( \frac{m^l t_\beta}{v}-\frac{\Sigma^l}{\sqrt{2} c_\beta} \Big) e_R\Big] A +\rm{H.C.}\nonumber
\\
\label{eq:yuk}
\end{eqnarray}
where $m^i$ corresponds to the diagonalized mass matrices of fermions, $s_\alpha=\sin\alpha$, $c_\alpha=\cos\alpha$, $t_\beta=\tan\beta$, $s_{\beta-\alpha}=\sin(\beta-\alpha)$ and $c_{\beta-\alpha}=\cos(\beta-\alpha)$. The $\Sigma$ matrices contain the off-diagonal entries and can induce tree-level FCNC. They are defined as $\Sigma^u= U^u_L Y^{u}_{1} {U^u_R}\dagger $, $\Sigma^d= U^d_L Y^d_{1} {U^d_R}^\dagger $ and $\Sigma^l= U^u_L Y^l_{2} {U^l_R}^\dagger $, $U$'s being the bi-unitary transformations required to diagonalize fermion mass matrices. In the $\Sigma^i \to 0$ limit, the Yukawa sector reduces to the same as pure Type-X 2HDM. $\Sigma^f$ can be parametrised as \cite{PhysRevD.35.3484}
\begin{equation}
    \Sigma^f_{ij}= \sqrt{m^f_i m^f_j} \frac{\chi_{ij}}{v} 
\end{equation}
For simplicity, we consider $\chi_{ij}$ to be symmetric. For three generations of leptons, the leptonic non-diagonal couplings ($\chi_{ij}$) would be noted as $y_{\mu e}$, $y_{\tau \mu}$ and $y_{\tau e}$ in the following sections. The constraints from lepton flavour violation put limit on these set of couplings as discussed in the next section. Here for our analysis, the leptonic couplings with CP odd scalar $A$ would be relevant.
\section{Muon (g-2) in connection to 2HDM and Lepton Flavor Violation}
\label{sec:muong-2}

Anomalous magnetic dipole moment of muon still continues to be in need of some explanation other than just SM.  The gyromagnetic ratio of the muon ($g_\mu$) is 2 at the classical level. However, it receives quantum corrections depending on the model considered and this correction is defined as $a_\mu = \frac{(g-2)_\mu}{2}$. The value of $a_{\mu}$ in the SM comes out to be~\cite{Aoyama:2020ynm,Aoyama:2012wk,Aoyama:2019ryr,Czarnecki:2002nt,Gnendiger:2013pva,Davier:2017zfy,Keshavarzi:2018mgv,Colangelo:2018mtw,Hoferichter:2019mqg,Davier:2019can,Keshavarzi:2019abf,Kurz:2014wya,Melnikov:2003xd,Masjuan:2017tvw,Colangelo:2017fiz,Hoferichter:2018kwz,Gerardin:2019vio,Bijnens:2019ghy,Colangelo:2019uex,Blum:2019ugy,Colangelo:2014qya}
\begin{equation}
 a^\text{SM}_{\mu} = 116591810(43) \times 10^{-11}. \label{eq:amuSM}
\end{equation}

 The "MUON G-2" collaboration at the Fermilab National Accelerator Laboratory (FNAL) in its recent report has published its latest experimental measurement of the anomalous magnetic moment of muon (muon $(g-2)$)~\cite{Muong-2:2023cdq,Aoyama:2020ynm}. The measurement at FNAL after improving the measurement uncertainty~\cite{Muong-2:2015xgu,Muong-2:2021vma,Muong-2:2021ojo} gives the value of the anomalous magnetic moment as~\cite{Muong-2:2021vma,Muong-2:2021ojo,Muong-2:2023cdq}
\begin{equation}
 a^\text{exp-FNAL}_{\mu} = 116592055(24) \times 10^{-11}.
 \label{Fermi}
\end{equation}

This new measurement from FNAL  along with a combination of old FNAL\,\cite{Muong-2:2021ojo,Muong-2:2023cdq} data and
older BNL(2006)\,\cite{Muong-2:2006rrc} data gives~\cite{Muong-2:2023cdq}
\begin{equation}
 a^\text{exp-comb}_{\mu} = 116592059(22) \times 10^{-11}
 \label{eq:expcomb}
\end{equation}
which results in an excess of $\Delta a_{\mu}= 249(48) \times 10^{-11}$.
 Although there are tensions in the Hadronic Vacuum Polarization (HVP)\,\cite{Kurz:2014wya,Davier:2017zfy,Keshavarzi:2018mgv,Colangelo:2018mtw,Hoferichter:2019mqg,Davier:2019can,Keshavarzi:2019abf} contribution to the $(g-2)$ due to the recent lattice QCD based results\,\cite{Blum:2019ugy,Borsanyi:2020mff,Ce:2022kxy,ExtendedTwistedMass:2022jpw,Chao:2022ycy} from BMW collaboration and the $e^+e^-\to \pi^+\pi^-$ data from CMD-3 experiment\,\cite{CMD-3:2023alj}. However, as any firm comparison of the muon (g-2) measurement with the theory is hard to establish, we, therefore, choose to work in the paradigm where a 5.1$\sigma$ excess exists, and a contribution from new physics is needed.

\begin{figure}
    \centering
    \includegraphics[scale=0.5
    ]{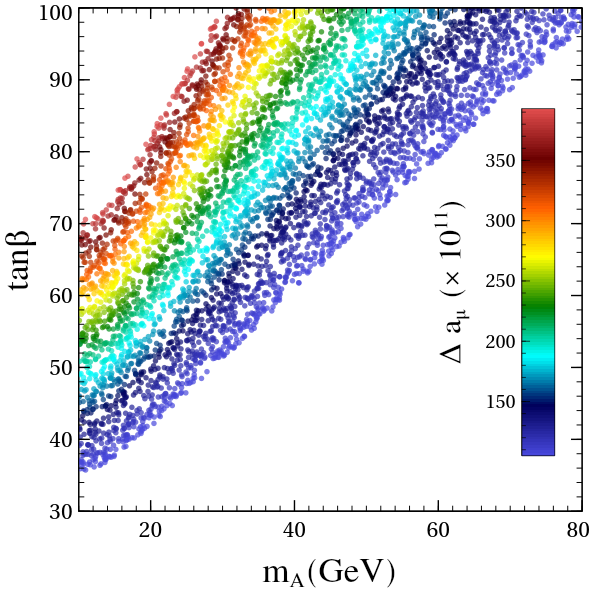}
    \caption{\it Parameter space allowed by muon anomaly data at 3$\sigma$ range in $m_A$-$\tan\beta$ plane. The mass of charged Higgs ($m_{H^+}$) and CP even Higgs ($m_{h}$) are taken to be 165 GeV and 110 GeV respectively. }
    \label{fig:amu}
\end{figure}

In this work, we take into account both the one-loop and two-loop Bar-Zee contribution to the muon anomaly in generalized 2HDM model~\cite{Ilisie:2015tra,Queiroz:2014zfa,Broggio:2014mna,Cherchiglia:2017uwv}. A detailed study in the context of $a_\mu$ has already been done in~\cite{Ghosh:2020tfq}. We scan our model parameter space imposing muon $g-2$ constraints and plot the allowed region in the $m_A$-$\tan\beta$ plane in Fig.\ref{fig:amu}. Here we see that points with low $m_A$ and large $\tan\beta$ are favored from muon anomaly data. While scanning, we have taken the 3$\sigma$ upper and lower bounds on observed
central value of $\Delta a_\mu$ as noted in Eq.~\ref{eq:expcomb}.

One must remember that the operators which contributes to $(g_{\mu}-2)$ are very similar to the ones which drive the flavour violating decay of charged leptons. It is well-known that non-observation of lepton flavour violating decays so far has severely constrained the respective branching ratios. ~\cite{BaBar:2009hkt}:
\begin{eqnarray}
    {\rm BR}(\mu \to e\gamma) &<& 4.2\times10^{-13}, ~~~~ {\rm BR} (\tau \to e\gamma) < 3.3\times10^{-8} \ ,\nonumber\\
    &&{\rm BR} (\tau \to \mu\gamma) < 4.4\times10^{-8}.
    \label{eq:LFV}
\end{eqnarray}

The strongest bound comes from the $ {\rm BR}(\mu \to e\gamma)$ process from MEG experiment~\cite{MEG:2016leq}. We find that to satisfy both the LFV and muon anomaly constraints, one needs to put the values of the $y_{e\mu}, y_{e\tau}$ and $y_{\mu\tau}$ respectively to be ${\cal{O}}(10^{-5}), {\cal{O}} (10^{-4})$ and ${\cal{O}}(10^{-5})$ or lesser. While scanning the parameter space for both the muon anomaly and LFV, we have chosen the other CP-even Higgs and the charged Higgs mass to be 110 GeV and 165 GeV respectively. These particular choices of the masses will be justified soon  in detail  in the following section. 

\section{Theoretical and experimental constraints on model parameters:} \label{sec:cons}
Before going into the details of $A$ production at a muon collider, let us briefly discuss the relevant theoretical and experimental constraints on the model parameters. For scanning of the parameter space, we have considered the alignment limit in our analysis and therefore have kept the mixing angle $\cos(\beta - \alpha)$ to be close to unity. The scan ranges of the parameters are mentioned below:
\begin{eqnarray}
m_{12}^2&\in&[-500,500]~{\rm GeV^2},
m_A \in [10.0, 60.0]~{\rm GeV},  \nonumber \\
m_H &\in& [62.5, 125.0]~{\rm GeV}, m_H^{\pm}\in [89.0, 190.0]~{\rm GeV}, \nonumber \\
\tan \beta &\in& [10, 100],~~~~|\cos(\beta - \alpha)| \in [0.99, 1],\nonumber \\
\lambda_6 &\in & [0, 0.1],~~~\lambda_7 \in [0, 0.1] 
 \label{scan}
\end{eqnarray}

$\bullet$ ~~\textbf{Vacuum Stability and Perturbativity Unitarity:}
The necessity to obtain a stable vacuum imposes following constraints on the Higgs quartic couplings. The set of stability conditions for this model are as follows 
\begin{eqnarray}
    \lambda_1 >0, ~~ \lambda_2 > 0, ~~ \lambda_3 > -\sqrt{\lambda_1 \lambda_2},~~ \lambda_3 +\lambda_4-\lambda_5 > \sqrt{\lambda_1 \lambda_2}
\end{eqnarray}
On the other hand, unitarity demands the $\lambda$ parameters to be less than $\sim 4\pi$.
The quartic couplings can be expressed in terms of the physical parameters such as the mass of the particle, vev  etc and therefore we can translate these bounds to the physical parameter spaces. The other crucial parameter for perturbative unitarity is the soft $Z_2$ breaking parameters which requires to be $m^2_{12} \simeq \frac{m^2_H}{\tan\beta}$ to ensure $\lambda_1$ to be within the perturbative limit~\cite{Ghosh:2020tfq}. These conditions for vacuum stability and unitarity of 2HDM have been previously discussed in multiple works \cite{Broggio:2014mna,PhysRevD.67.075019}.  As shown in Ref. \cite{Ghosh:2020tfq}, though low to moderate $\tan\beta$ values are preferred to satisfy the abovementioned constraints for  $m_A$ ranging between ($10- 60$~GeV), but higher values of $\tan\beta$ can also satisfy the constraints for relatively lower number of parameter points. Here for our choice of CP odd mass (in the range $30-50$ GeV),  higher $\tan\beta$ values are preferred in order to satisfy the muon $g-2$ constraint in the 3$\sigma$ limit. We scan our parameter space for low $m_A$ and high $\tan\beta$ using  {\tt 2HDMC-1.8.0} \cite{Eriksson:2009ws} package and have found points where vacuum stability, unitarity, and perturbativity constraints are satisfied.  \\ \\

$\bullet$ ~~\textbf{Electroweak constraints}

\noindent Due to the presence of non-standard Higgs in the current scenario, the $W$ and $Z$ boson receive one-loop correction to their masses and therefore the oblique parameters $S, ~T, ~U$~\cite{Peskin:1990zt,Peskin:1991sw} are modified. Consideration of updated values of SM Higgs mass and top mass gives the following values of $S, ~T, ~U$ \cite{Haller:2018nnx}
\begin{equation}
    S= 0.04\pm 0.11, ~ T=0.09\pm 0.14, ~U=-0.02\pm 0.11
\end{equation}
\noindent This in turn restricts the mass gap between the charged and the light CP even Higgs. For our choice of benchmark points, where this mass difference is $-55$ GeV, the electroweak observables are within the 2$\sigma$ allowed range.  \\ \\

$\bullet ~~\textbf{B-Physics constraints:}$

The presence of the flavor-changing terms in
the Yukawa Lagrangian of the charged Higgs (Eq.~\ref{eq:yuk}) leads to rare processes involving B-mesons~\cite{Crivellin:2013wna,Arbey:2017gmh,Hussain:2017tdf}. In the presence of non-zero FCNC Yukawa matrix elements, the $B \to X_s \gamma$ process will be modified. However, even in this scenario, it is possible to have low enough charged Higgs mass $m_{H^{\pm}} \gtrsim 150$ GeV by taking $\lambda_{tt} \sim 0.5$ and $\lambda_{bb} \sim 2$~\cite{Xiao:2003ya,Mahmoudi:2009zx,Arhrib:2017yby,Cherchiglia:2017uwv,Enomoto:2015wbn}. 
The other decay process which can constrain our model parameters space is $B^{\pm} \rightarrow \tau^{\pm} \nu_{\tau}$ where charged Higgs enters at the tree level itself~\cite{Alonso:2016oyd}. The constraint from $\Delta M_B$ also puts an an upper limit on $\lambda_{tt}$ as a function of the charged Higgs mass~\cite{Mahmoudi:2009zx}. $m_{H^{\pm}} \gtrsim 150$ GeV is allowed for $\lambda_{tt} \lesssim 0.5$. The upper limit on the BR($B_s \rightarrow \mu^+ \mu^-$)~\cite{ParticleDataGroup:2022pth} constrains the low $\tan \beta (< 2$) region for low $m_H^{\pm}(\sim 100$ GeV)~\cite{Arbey:2017gmh}. For higher charged Higgs mass this limit is further relaxed. Therefore, these specific searches do not significantly impact our parameter space. \\ \\

$\bullet ~~\textbf{Constraints from collider searches:}$

2 HDM, being a very well-explored choice of model, faces plethora of constraints from colliders. LEP experiment puts a tight bound on the charged Higgs mass from the $\tau\nu$ and $c\bar{s}$ channel to be $m_H^\pm >$ 80 GeV~\cite{ALEPH:2013htx}. At the LHC, an upper limit on the charged Higgs mass comes from the production ${\rm BR}(t \to b H^\pm)$ in the $\tau\nu$~\cite{ATLAS:2018gfm} and $c\bar{s}$~\cite{ATLAS:2013uxj} channels, when $m_h^\pm < m_t$. There are also available bounds on the charged Higgs mass from the search in $pp \to t b H^\pm$~\cite{ATLAS:2016avi,ATLAS-CONF-2016-088, Aaboud:2631950,sirunyan2020search,CMS-PAS-HIG-16-031,ATLAS-CONF-2016-089}.
We have taken into account all these searches and have set our charged Higgs mass to be 165 GeV.

The constraints coming from the direct search of the nonstandard neutral Higgs can also modify our parameter space. Specifically, as we are interested in the low pseudoscalar-mass region with enhanced coupling to leptons, the search for low-mass pseudoscalar produced in association with b quarks and decaying into a $\tau\tau$
final state~\cite{CMS:2015qnd,CMS:2019hvr} plays an important role. The search for low-mass (pseudo)scalars produced in association with $b\bar{b}$ and
decaying into $b\bar{b}$~\cite{CMS:2015grx,CMS-PAS-HIG-16-025} have been taken into account in our work.
CMS has also investigated decays involving two non-standard Higgs bosons, such as $h/H \to Z(\ell\ell)A(\tau\tau)$~\cite{CMS:2016xnc} and
$h/H \to Z(\ell\ell)A(b\bar{b})$~\cite{CMS-PAS-HIG-18-012,cms2019search}. However, these constraints are relevant only for heavier CP-even Higgs bosons with masses $\gtrsim$200 GeV.
Therefore, our parameter space is not affected by these constraints. 

CMS and ATLAS have a series of searches of the decay of 125 GeV Higgs at various final states, namely, $\tau\tau$~\cite{ATLAS:2015xst,ATLAS:2023qpu}, $\mu\mu$~\cite{ATLAS:2014hrp,CMS-PAS-HIG-19-006}, $ee$ {\cite{CMS-PAS-HIG-21-015} and also lepton flavor violating $e\tau$~\cite{ATLAS:2023mvd}, $\mu\tau$~\cite{ATLAS:2023mvd}, and $\tau\tau$~\cite{CMS:2019pex} channels. However, our choice of $\cos(\beta-\alpha)\simeq 0.99$ and higher CP-even Higgs to be 125 GeV, helps us to satisfy the lepton-flavor-violating decays of
the 125 GeV Higgs trivially, as the coupling goes as $\sin^2(\beta-\alpha)$ as we see from Eq.~\ref{eq:yuk}.

The most stringent condition that constrains our model parameter space is the 125 GeV Higgs decaying to a pair of light pseudoscalars~\cite{CMS:2020ffa,CMS:2024zfv}. In our work, we have taken the higher CP-even Higgs $m_H$ to be 125 GeV. However, in this case, from consideration of the LEP bounds, we know that either $m_A$ or $m_h$ can be $< m_H/2$. As we are interested in low mass pseudoscalar for the collider analysis, we keep $m_h=$ 110 GeV, i.e. $> {m_h}/2$. For detailed discussion, please see Refs.~\cite{Ghosh:2020tfq,Ghosh:2021jeg}. We have explicitly checked that $HAA$ coupling (Eq.~\ref{eq:haa}) can be made very small by suitable adjustment of $\tan\beta, m^2_{12}$ and $m_h$, thus avoiding the direct search constraints from $H\to A A$.
\begin{eqnarray}
g_{HAA}&=& \frac{1}{2v}[(2m_A^2 - m_H^2) \frac{\cos(\alpha-3\beta)}{\sin 2\beta} + (8 m_{12}^2 \nonumber\\
&&-\sin 2\beta (2m_A^2 + 3m_H^2)) \frac{\cos(\beta+\alpha)}{\sin^2 2\beta}]\nonumber \\
&+& v\Big[\sin 2\beta \cos 2\beta(\lambda_6-\lambda_7)\cos(\beta - \alpha) \nonumber \\        &+& (\lambda_6 \sin\beta \sin 3\beta + \lambda_7 \cos \beta \cos 3\beta)\sin(\beta - \alpha)\Big]\nonumber \\ 
\label{eq:haa}
\end{eqnarray}

We conclude this section with the remark that we have taken $m_h = 110$ GeV, $m_H=125$ GeV, $m_H^\pm=165$ GeV, $m_A \in [30,50]$ GeV and $\tan\beta \in [50,80]$ for our collider analysis.   
\begin{table*}[tbh]
\centering\resizebox{15cm}{!}{
\begin{tabular}{llllllllll}
  \toprule \hline
 BP & $\tan\beta$ & $m_{A}$(GeV) & $m_{h}$(GeV) & $m_{H^{\pm}}$(GeV) & $m^2_{12}(GeV^2$)& $\lambda_6$ & $\lambda_7$ & $|s_{\beta\alpha}|$ & $\sigma_{prod}(fb)$ \\
  \midrule
  BP1 & 54.96 & 30 & 110 & 165 & 284.27 & 0.004 & 0.0004 & 0.01 & 0.43\\
 BP2 & 58.94 & 35 & 110 & 165 & 265.08 & 0.003 & 0.0007 & 0.01 & 0.49 \\
 BP3 & 68.84 & 40 & 110 & 165 & 226.97 & 0.01 & 0.0006 & 0.01 & 0.67\\
 BP4 & 74.66 & 50 & 110 & 165 & 209.27 & 0.004 & 0.0005 & 0.01 & 0.78 \\
  \bottomrule
\end{tabular}}
\caption{Benchmark points allowed by all theoretical and experimental constraints and effective cross-section for $\gamma\tau^+\tau^-$ channel at 3 TeV muon collider.}
\label{tab:cross-sec}
\end{table*}

\section{Collider Searches}
\label{sec:coll}
In this section, we try to probe the pseudoscalar-muon coupling, one of the controlling factors of muon $g-2$, through the search of pseudoscalar at a muon collider.
Here, we explore the production of a mono-photon in association with a CP odd scalar $A$ that further decays into two $\tau$'s in lepton-specific 2HDM at muon collider as:
 \begin{equation}
     \mu^+ \mu^- \to \gamma \, A \to \gamma \, \tau^+ \tau^-,
\label{eq:coll}      \end{equation}
 followed by the leptonic decay of the $\tau$'s. In Fig.~\ref{fig:feynmancoll}\,\footnote{Here we would like to mention that we have also considered the contribution coming from the s channel and have generated the process mentioned in Eq.~\ref{eq:coll} inclusively. However, for the region of parameter space considered and the signal topology, the s- channel contribution with respect to the t- channel diagrams is negligible. Another important point to mention here is that a diagram with s- channel in one loop via $A-\gamma-\gamma/Z$ effective coupling can contribute in this process. However this contribution is atleast $\alpha^2 (m_{\tau}/m_{\mu})^2 $ times suppressed which is less than $1\%$ of the dominant tree-level contribution. Therefore, we have safely neglected these contributions in the collider analysis.}, we show the Feynman diagram which dominantly contributes to the process  mentioned in Eq.~\ref{eq:coll}. Therefore, our signal of interest is $\ell^+\ell^{'-}\gamma+\mET$ where $\ell,\ell^\prime = e , \mu$. The SM processes that can mimic this signal are $\gamma \, W^+ W^-, \gamma \, Z Z$ and $\gamma \, \tau^+ \tau^-$. The first two among the aforementioned processes are the dominant backgrounds in our signal region. We have explicitly checked that the third background can be reduced completely by applying a cut on the separation of the two leptons ($\Delta R_{l l^\prime}$) which will be discussed shortly. Therefore we do not discuss the $\gamma \, \tau^+ \tau^-$ background here in detail. 

 \begin{figure}[H]
    \centering
    \includegraphics[scale=0.1]{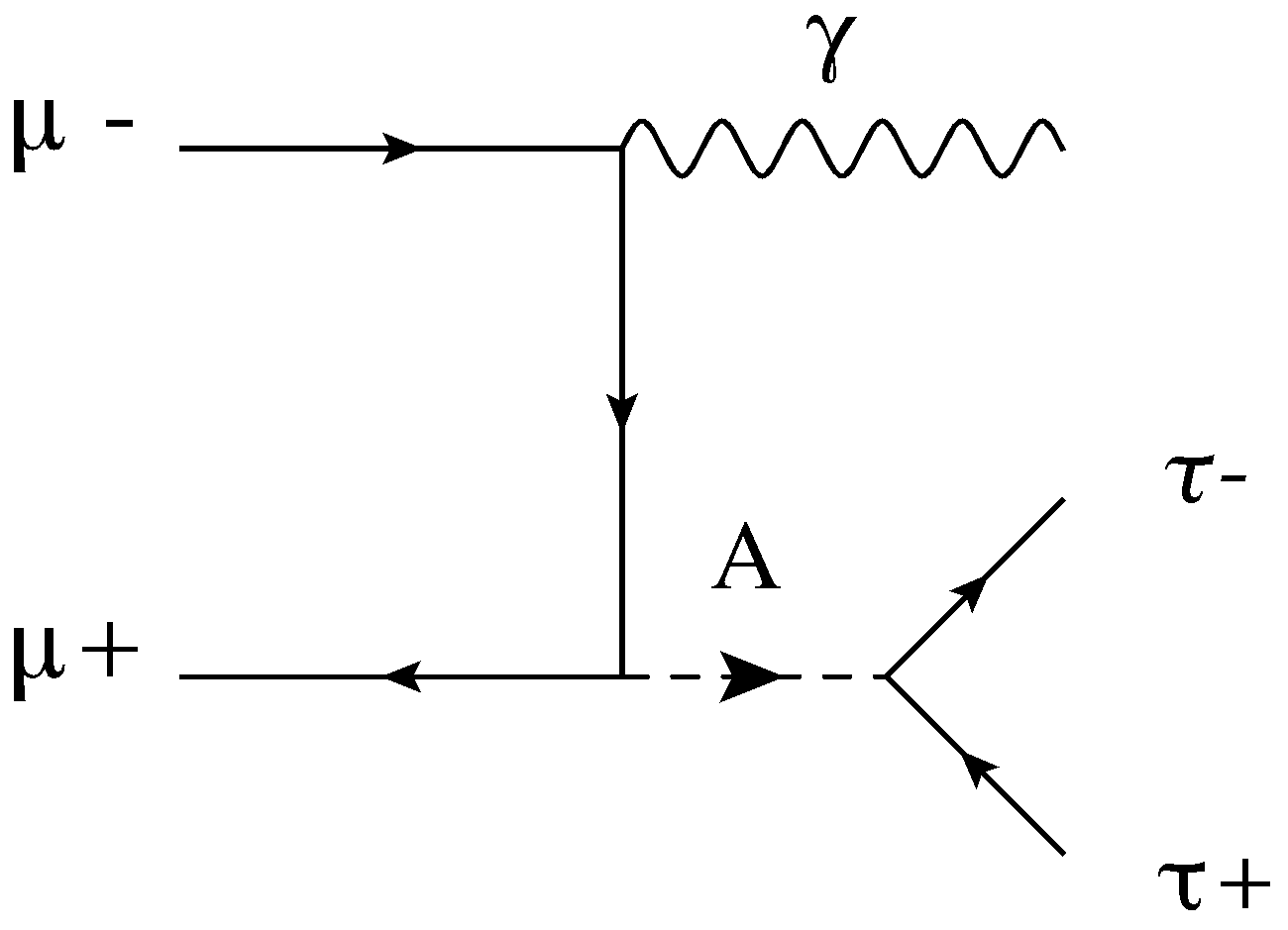}
    \caption{Feynman diagram with the dominant contribution for the signal process $\mu^{+} \mu^{-} \to \gamma \,\tau^{+}\tau^{-}$ }
    \label{fig:feynmancoll}
\end{figure}

We analyze four benchmark points that satisfy all necessary theoretical conditions (vacuum stability, unitarity) and experimental constraints such as constraints from oblique parameters, muon anomaly in $3\sigma$ limit   and lepton flavor violation constraints. The choice of parameters for the benchmarks and corresponding cross-sections are tabulated in Table~\ref{tab:cross-sec}. Here we see that the cross section of the process strongly depends on $\tan\beta$ which is the coupling of the pseudoscalar with muon apart from the $m/v$ factor. As the mass of the pseudo scalar is considered to be very light (around 30 $\sim$ 50 GeV) in comparison to the centre of mass energy (3 TeV) in our analysis, the cross-section does not depend on the pseudo-scalar mass. Therefore, the cross-section is solely dictated by the pesudo-scalar-muon coupling which controls the contribution in muon $g-2$. In the following, we present a cut-based analysis of the above-mentioned channel to probe this coupling. 

To analyse the collider aspects, we have implemented the model in {\tt FEYNRULES}\cite{Alloul:2013bka} and generated the UFO file. The signal and background generation is done by feeding the UFO file in {\tt MadGraph5@NLO}\cite{Alwall:2011uj}. {\tt PYTHIA8}\cite{Sjostrand:2014zea} is used for hadronization and showering. The showered events are then passed through {\tt DELPHES}\cite{deFavereau:2013fsa} for detector simulation purposes with the necessary modified muon collider card~\cite{muoncard}.                   
The preselection cuts used to generate the background and events are as follows 
\begin{equation}
   |\eta(\gamma)| < 2.5 ; ~~|\eta(l)| <2.5 ; ~~p_T(l) > 10~{\rm GeV} ; ~~p_T(\gamma) > 10~{\rm GeV};   
\end{equation}
In addition to the basic cuts, we propose this set of selection cuts over the following kinematic observables which reduce the SM background events and improve the signal significance significantly.

\begin{figure*}[tbh!]
   \centering
   \subfigure[]{ \includegraphics[scale=0.2]{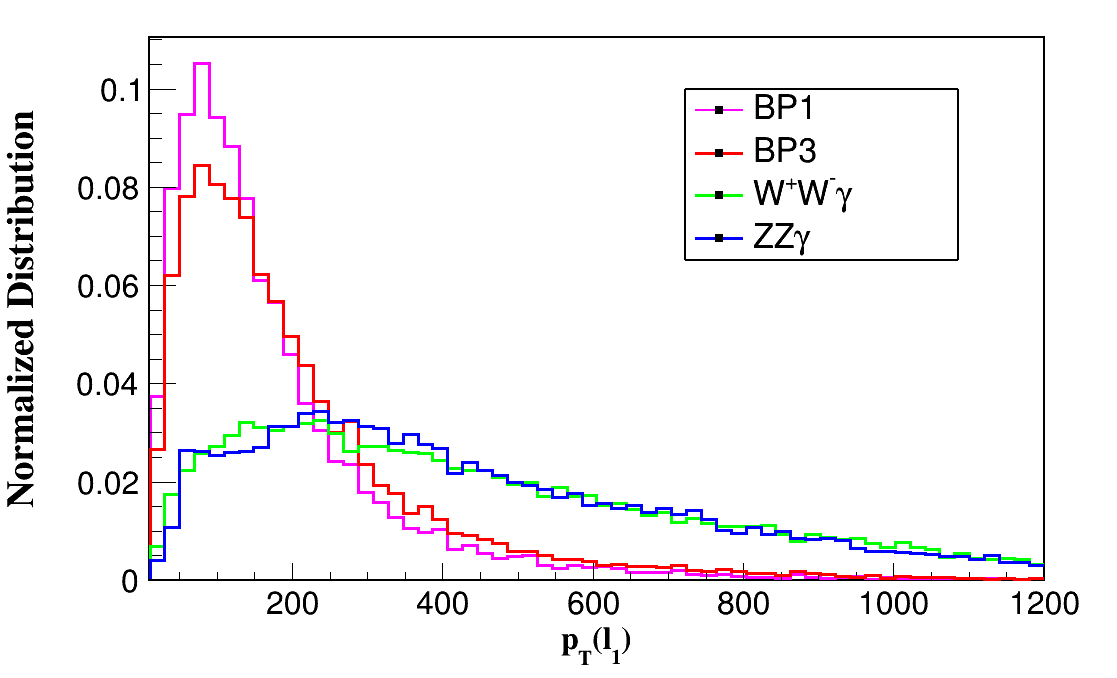} }~  
   \subfigure[]{  \includegraphics[scale=0.2]{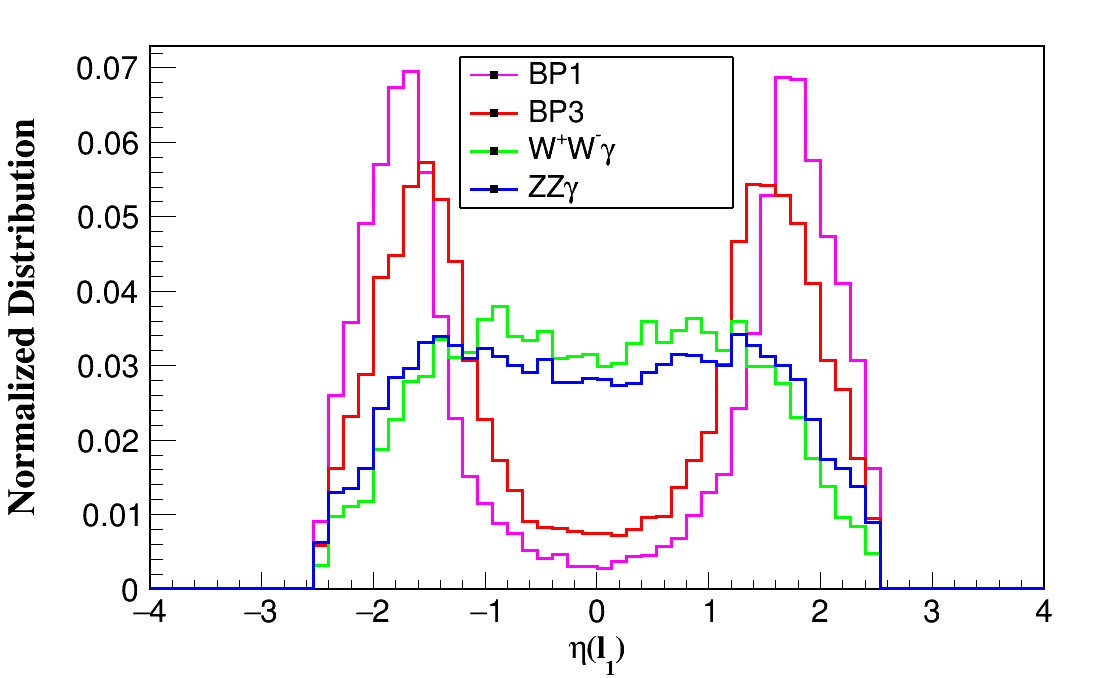}}
  \\
   \subfigure[]{ \includegraphics[scale=0.2]{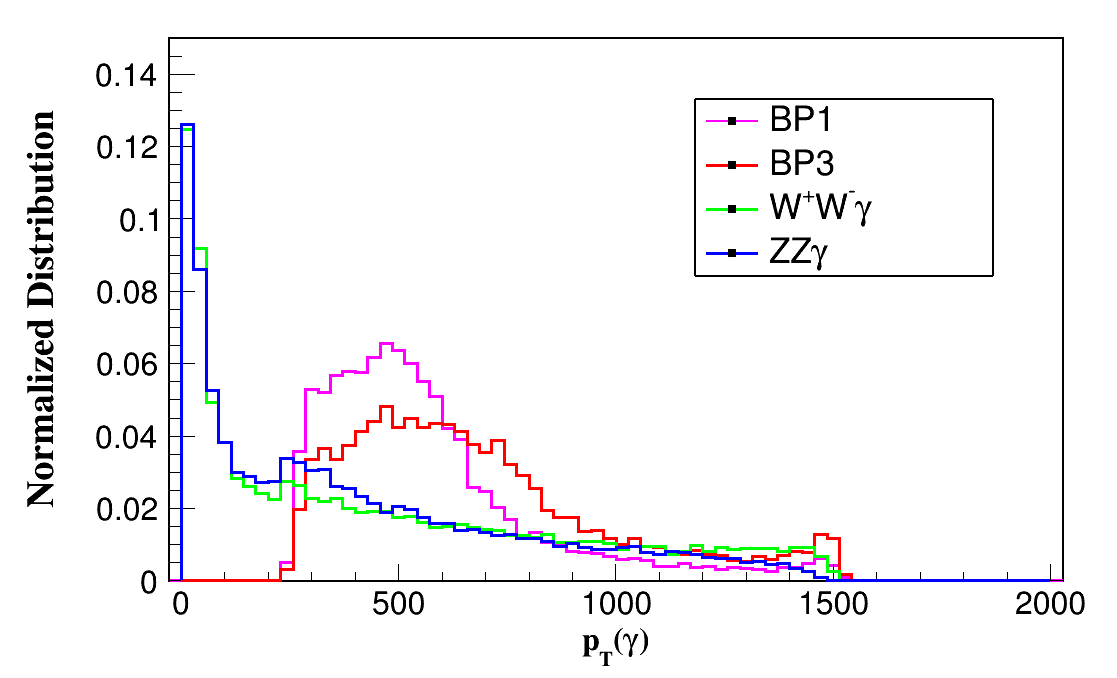}}~
    \subfigure[]{ \includegraphics[scale=0.2]{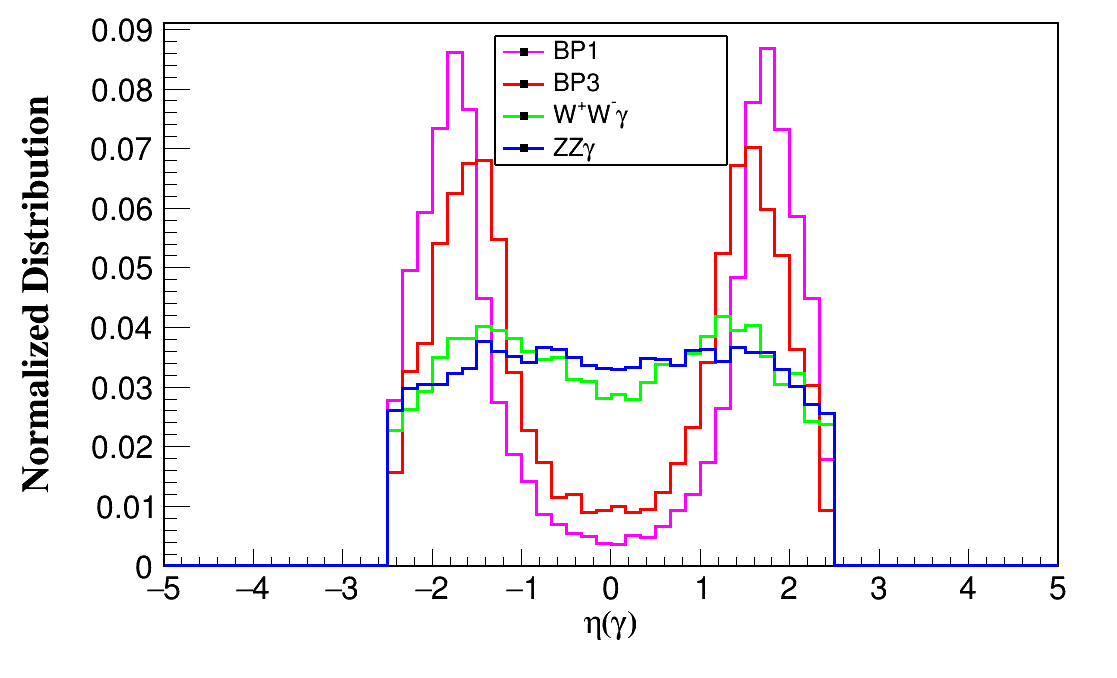}} 
    \\
  \subfigure[]{  \includegraphics[scale=0.2]{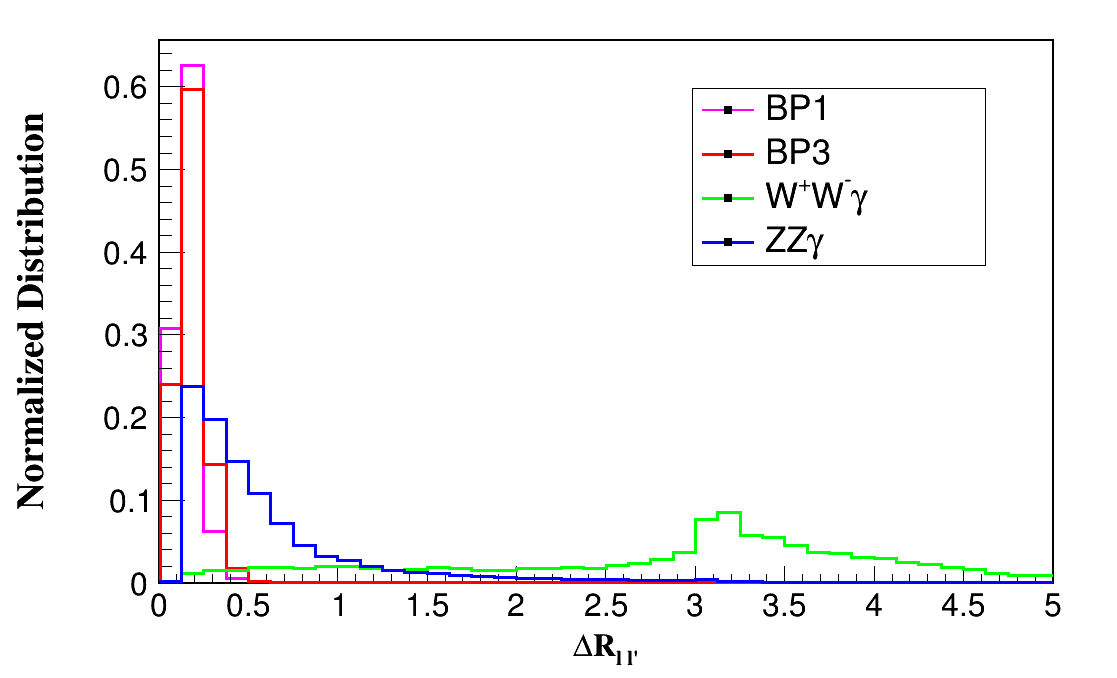}}~
    \subfigure[]{ \includegraphics[scale=0.2]{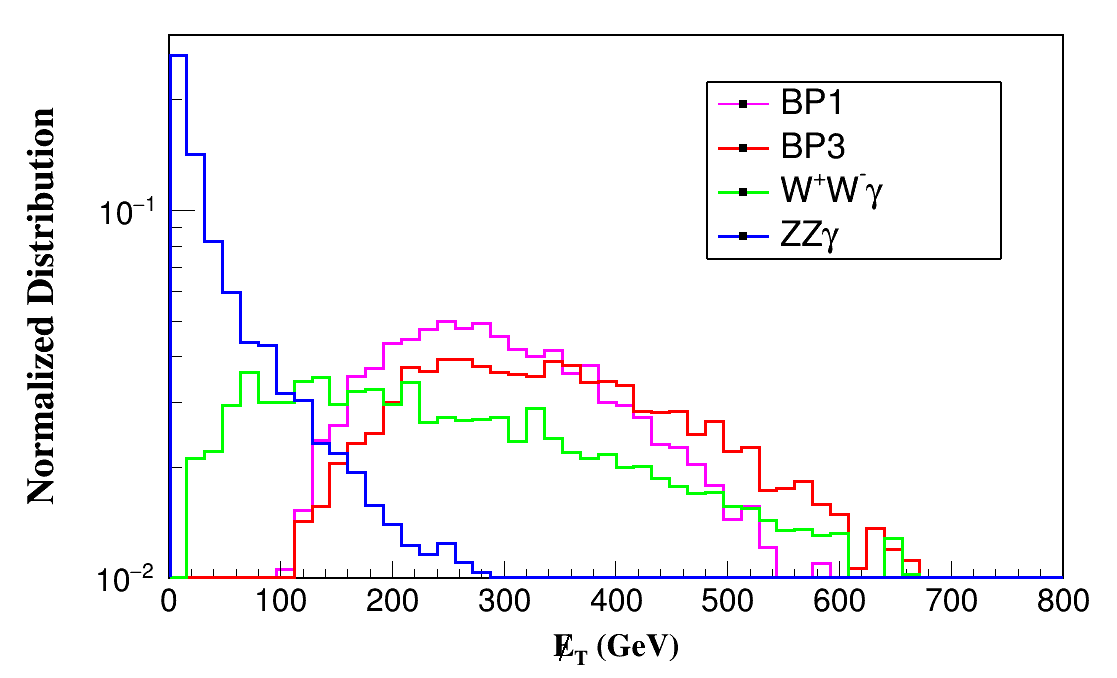}}

    \caption{Normalized distribution of different kinematic variables for both the signal and background events. }
        \label{fig:kinematicvariable}
\end{figure*}

$\bullet$ \textbf{$p_T$ of the leptons ($p_T(\ell)$)}:  In left panel of Fig.~\ref{fig:kinematicvariable}(a), we show the transverse momentum ($p_T$ in GeV) distribution of the leading lepton. For the signal, the leptons come from the decays of two $\tau$'s which are products of the pseudoscalar. As a result, the leptons are peaked at lower values of $p_T$. For the backgrounds, the leptons come from the boosted $W^\pm$ and $Z$ bosons which results in peaking at relatively higher values. Therefore applying a cut over $p_T <$ 300 GeV reduces the background considerably.

$\bullet$ \textbf{$\eta$ of the lepton ($\eta_l$)}: We show the $\eta$ distribution of the leading lepton in Fig.~\ref{fig:kinematicvariable} (b). Due to the high boost of the signal leptons, the rapidity peaks around the higher value region (close to the beam axis) whereas the background is almost uniformly distributed from $-2.5$ to $2.5$ in the $\eta_l$ axis. 
Background events are reduced by demanding $|\eta_{\ell_1}| > 1.5$ for BP1 and BP2 where as $|\eta_{\ell_1}| > 1.0$ has been used for BP3 and BP4.

$\bullet$ \textbf{$p_T$ of the photon ($p_T(\gamma)$)}: In Fig.\,\ref{fig:kinematicvariable} (c), we show the momentum distribution of the photon. For the signal process, the photons come directly from the muons. As a result, they tend to peak at higher momentum. For the background processes, due to 3-body decay, the low momentum sides are mostly populated with a long tail. 
We see that choosing $p_T(\gamma) >$ 200 GeV
greatly reduce background in comparison to signal events.  

$\bullet$ \textbf{$\eta$ of the photon ($\eta_\gamma$)}: We portray the rapidity distributin of photon in Fig.\,\ref{fig:kinematicvariable} (d). This is very similar to the lepton rapidity distribution. Benchmark-specific cuts can reduce backgrounds without affecting the signal events.   

$\bullet$ \textbf{$\Delta R$ between the leptons ($\Delta R_{l l^\prime}$)}: In Fig.~\ref{fig:kinematicvariable} (e), $\Delta R_{l l^\prime}$ distribution is shown for the signal and background events. For $W^+ W^-$ background, the leptons coming from two different particles, acquire a bigger cone and, therefore are distinguishably distant from the signal distribution as shown in the plot. However, for the $ZZ$ background, the distinction is difficult as the two leptons originates from one mother particle for both cases. But the CP odd scalar $A$ being lighter than the $Z$ boson, the signal distribution peaks towards the lower end of the $\Delta R$ axis more in comparison to the $ZZ$ background. Therefore, an appropriate cut of $\Delta R <$ 0.35 takes the main role in increasing the signal significance.  

$\bullet$ \textbf{ missing transverse energy ($\mET$)}:
The $\mET$ appears from the neutrinos for both the signal and background events. Though the $ZZ$ background can greatly be reduced by applying a cut over missing $\mET$ as portrayed in the right panel of Fig.\,\ref{fig:kinematicvariable} (f), however, the main background which arises from $WW$ process, can not be reduced by applying this cut. To ensure that our signal contains $\mET$, we put a basic cut of $\mET > 10$ GeV while generating the events and refrain from applying any further hard cut on this variable.  

After applying proper cuts on the aforementioned observables, signal significance has been calculated in Table~\ref{tab:signi} using the following formula \cite{Cowan:2010js}
\begin{equation*}
    \mathcal{S}=\sqrt{2[(S+B){\, \rm ln \,}(1+\frac{S}{B})-S]}.
\end{equation*}
where S (B) is the number of signal events (background events) after applying all cuts. In Table~\ref{tab:signi}, we see that all four benchmark points can be probed with significance $\gtrsim$ 4$\sigma$ with 1 $\rm ab^{-1}$ luminosity at the proposed 3 TeV muon collider.

\begin{table*}
	\centering
 \resizebox{15cm}{!}{
	\begin{tabular}{c c c c c c c}
		\cline{1-7}
		 \multicolumn{7}{c}{Number of events after cut ($\mathcal{L}=1 ab^{-1}$)}  \\ \cline{1-7}
		SM-background  
		& Preselection cuts & $p_T(\ell_{1})< 300$ GeV  & $|\eta_{l_1}| >$ 1.5 & $p_T (\gamma)>200$ GeV & $|\eta_\gamma|>$ 1.5  & $\Delta R_{l l^\prime} < 0.35$   
		\\ 
                  ~$\gamma ~W^+ W^-$ & 966 & 369 & 171 & 123 & 98 & 2 
		\\
  
                  $\gamma Z Z$ & 136 & 51 & 27 & 15 & 7 & 0   

		\\
		Signal  
		&  &   & &  &   &    
		\\ 
		\multicolumn{1}{c}{BP1}  & 12 & 10 & 8 & 8 & 8 & 8 \\

		\multicolumn{1}{c}{BP2}&  17  & 14 & 10 & 10 & 10 & 10 \\ \hline
  
\end{tabular}}

\vspace{0.5cm}
\centering
	\resizebox{15cm}{!}{
	\begin{tabular}{c c c c c c c}
		\cline{1-7}
		 \multicolumn{7}{c}{Number of events after cut ($\mathcal{L}=1 ab^{-1}$)}  \\ \cline{1-7}
		SM-background  
		& Preselection cuts & $p_T(\ell_{1})< 300$ GeV  & $|\eta_{l_1}|>$ 1.0 & $p_T (\gamma)>200$ GeV & $|\eta_\gamma|>$ 1.0  & $\Delta R_{l l^\prime} < 0.35$   
		\\ \cline{1-7} 
                  ~$\gamma ~W^+ W^-$ & 966 & 369 & 260 & 187 & 172 & 5 
		\\

                  $\gamma Z Z$ & 136 & 51 & 36 & 24 & 16 & 0   

		\\
  
		Signal  
		&  &  &  &  &   &  
		\\ 
		\multicolumn{1}{c}{BP3}& 27 & 22 & 20 & 20 & 20 & 19 \\ 
  
                     \multicolumn{1}{c}{BP4}& 40 & 30 & 25 & 25 & 25 & 22\\ \hline

\end{tabular}}

\vspace{0.5cm}
	\begin{tabular}{c c}

 \hline \cline{1-2} \hline
 \cline{1-2}
 Signal & Significance(S) \\
 \hline \cline{1-2} \hline
 
  BP1 & 4.0 $\sigma$ \\
 BP2 & 4.8 $\sigma$ \\
 BP3 & 6.1 $\sigma$ \\
 BP4 & 6.9 $\sigma$ \\
 \hline \hline 
 \end{tabular}
 \caption{The cutflow for the signal and backgrounds for $\ell^+\ell^{'-}\gamma+\mET$ channel at proposed 3 TeV muon collider and the significance reach for the four benchmark points at 1$ab^{-1}$ luminosity.}
\label{tab:signi}
\end{table*}

Before concluding this section, we would like to comment regarding the possibility of probing the four benchmark points at 14 TeV HL-LHC. The cross-sections for this channel are  $1.8\times 10^{-3}$ fb and $8.9\times10^{-4}$ fb for BP1 and BP4 respectively, which are at least a factor of ${\cal{O}}(200(800))$ less than that of the cross-section compared to the 3 TeV muon collider as can be seen from Table~\ref{tab:cross-sec}. The reason behind such a small cross-section at HL-LHC is the fact that the quark to pseudoscalar coupling ($A$-$q$-$\bar{q}$) is proportional to $\cot\beta$, whereas the muon to pseudoscalar coupling ($A$-$\mu^+$-$\mu^-$ ) is proportional to $\tan\beta$ (Eq.~\ref{eq:yuk}). From muon $g-2$ data, we see that low $m_A$ and high $\tan\beta$ are preferred which in turn makes the search for this $\ell^+\ell^{'-}\gamma+\mET$ signal topology at muon collider much more lucrative than the LHC.
\section{Conclusion}
\label{sec:conc}
To summarise, we have investigated the prospect of probing the coupling of a light pseudo-scalar, generally present in 2HDM, to a pair of muons at a multi TeV muon collider facility. The importance of probing such a coupling lies in the fact that it plays a crucial roles in a model with light pseudo-scalar explaining $(g-2)_\mu$ anomaly.  We have identified the model parameter space where muon $g-2$ constraint is satisfied at $3 \sigma$ limit. The effect of theoretical constraints pertaining to the requirements of perturbativity, unitarity and vacuum stability are discussed on the model parameters allowed by muon anomaly data.  As our model contains non-diagonal Yukawa coupling, we have considered the B-physics constraints as well. Finally, we have taken into account direct searches for the SM Higgs along with the other scalars present in the model. As the main contribution to muon anomaly comes from the low-mass pseudoscalar, we have ensured that the branching fraction of SM Higgs to the pseudoscalar does not exceed the upper bound obtained from collider data by demanding that the observed
125 GeV Higgs is the heavier of the two CP-even states of the 2HDM in the alignment
limit.

After satisfying the theoretical and experimental constraints, we looked for a signal of the pseudoscalar at a 3 TeV muon collider. As the pseudoscalar has Yukawa couplings that are lepton(muon)-philic, this gives us a unique opportunity to look for a distinctive signal of $\ell^+\ell^{'-}\gamma+\mET$ channel. The main motivation for the search at the muon collider lies in the fact that this channel would have a smaller cross-section to be probed even at HL-LHC due to the suppressed coupling of the pseudoscalar with quarks for the parameter space favored by muon ($g-2$) anomaly data. The other advantage at a muon collider would be the absence of QCD backgrounds. After a simple cut-based analysis, we have shown that the pseudoscalar having a mass range of 30 to 50 GeV can be probed with significance $\gtrsim$ 4$\sigma$. As the luminosity reach of the muon collider is yet to be finalized, one can hope that even more parameter space can be probed at the muon collider in this signal topology with high luminosity.

\section*{Acknowledgement}
NG is thankful to IISc for financial support from the IOE postdoctoral fellowship. ND and NG thank Dr. Tousik Samui for the creative discussion. The authors would like to thank Prof. Dilip Kumar Ghosh and Prof. Sunanda Banerjee for their useful discussions. ND would like to thank Prof. Anindya Datta for useful comments.  NG would also like to acknowledge Dr. Jayita Lahiri for informative conversations. ND is funded by CSIR, Government of India, under the NET SRF fellowship scheme with Award file No. 09/080(1187)/2021-EMR-I.

\bibliography{ref.bib}
\end{document}